\documentstyle[aps,twocolumn,epsfig]{revtex}

\draft

\begin{document}

\title{Collective oscillations of a classical gas confined in harmonic
traps}

\author{David Gu\'ery-Odelin$^{1,2}$, Francesca Zambelli$^1$,
 Jean Dalibard$^{2}$, Sandro Stringari$^{1}$}

\address{$^1$Dipartimento di Fisica, Universit\`a di Trento,}

\address{and Istituto Nazionale per la Fisica della Materia, I-38050
Povo, Italy}

\address{$^2$Laboratoire Kastler Brossel, Ecole normale sup\'erieure}
\address{24, Rue Lhomond, F-75231 Paris Cedex 05, France}

\date{\today}

\maketitle

\begin{abstract}
\noindent
Starting from the Boltzmann equation we calculate the frequency and the 
damping of the monopole and quadrupole oscillations of a classical gas 
confined in an harmonic potential. The collisional term
is treated in the relaxation time approximation and a gaussian ansatz is 
used for its evaluation.
Our approach provides an explicit description of the transition between 
the hydrodynamic and collisionless regimes in both spherical and deformed 
traps. The predictions are compared with the results of a numerical 
simulation.
\end{abstract}


\narrowtext

\begin{section}{Introduction}

After the experimental realization of Bose-Einstein condensation in
trapped atomic gases \cite{bec}, the investigation 
of collective excitations in these systems has become a very popular
subject of research (see \cite{rmp} for a recent theoretical review). At very
low temperature, when the the whole system is Bose-Einstein condensed, the
motion is described by the hydrodynamic equations of superfluids.
These equations,  which can be directly derived starting from the mean 
field Gross-Pitaevskii equation for the order parameter, give
predictions \cite{hds} in very good agreement with  experiments
 \cite{exp}. At higher temperatures the  mean field effects become
less important, while collisional terms cannot be longer ignored. 
If the temperature is  notably larger
 than the critical temperature for Bose-Einstein condensation the 
dynamical behaviour of a dilute gas is well described by the Boltzmann 
equation. 
In this case  two different regimes may occur: 
a collisional (hydrodynamic) regime characterized
by conditions of local statistical equilibrium and a collisionless regime
where
the motion is described by the single particle hamiltonian. 
Differently from the case of uniform gases, also in the
collisionless regime the system exhibits  well defined oscillations which
are driven by the external confinement. 
The equations for the collisional regime were
investigated in  \cite{Griffin,Kagan}, while a phenomenological
interpolation between the two regimes was proposed in \cite{Pethick}. 

The purpose of this paper is to provide  an analysis of the lowest
oscillation modes (eigenfrequency and damping) using the 
classical Boltzmann equation. In particular we study
the transition between the 
hydrodynamic and collisionless regimes. Our approach is based
on the relaxation time approximation for the collisional integral. The
corresponding
relaxation time is explicitly calculated 
by means of a gaussian ansatz for the distibution function which allows 
for an analytic evaluation of the collisional contribution to the
equations of motion.

For harmonic trapping the 
gaussian ansatz exactly reproduces the solution of the classical
Boltzmann equation in both the collisional and collisionless regimes and 
is consequently expected to be a good approximation also in the 
intermediate regime. This is explicitly checked by comparing the 
analytic predictions of the gaussian ansatz with the exact
results of a numerical simulation based on molecular dynamics.

In our paper atoms behave like hardspheres, $\sigma_0$ being their total 
cross section which will be assumed to be energy independent.
This is well satisfied in classical ultracold gases
where  collisions are completely characterized by the {\it s}-wave
scattering 
length and the cross section is thus isotropic and in most cases energy 
independent.

\end{section}

\begin{section}{Method of averages}

The starting point of our analysis  is the Boltzmann equation
for the phase space distribution function $f({\bf r},{\bf v},t)$ 
\cite{Huang}:
\begin{equation}
\frac{\partial f}{\partial t}+{\bf v}\cdot\mbox{\boldmath $\nabla$}_{\bf
r}f+\frac{{\bf F}}{m}\cdot\mbox{\boldmath $\nabla$}_{\bf v}f=I_{\rm
coll}(f)
\label{Boltzmann}
\end{equation}
where 
\begin{eqnarray}
I_{\rm coll}(f(\bf r,\bf v_1))&=& \frac{\sigma_0}{4\pi}\int d^2\Omega\; 
d^3{v}_2\;|{\bf v}_2-{\bf v}_1|
\nonumber\\
&&\big[f({\bf v}_1^{\prime})f({\bf v}_2^{\prime})-f({\bf v}_1)f({\bf
v}_2)\big]
\nonumber
\end{eqnarray}
is the usual classical collisional integral.

It accounts for elastic collisions between particles $1$ and $2$, 
with initial velocities ${\bf v}_1$ et ${\bf v}_2$, and final
velocities 
${\bf v}_1^{\prime}$ and ${\bf v}_2^{\prime}$. The solid angle $\Omega$ 
gives the direction of the final relative velocity.
The expression for the collisional 
term can be easily extended to include effects of both Bose and Fermi 
statistics \cite{baym}.
Actually most of the results discussed in this paper hold also in the
presence
of quantum degeneracy, provided the system is not Bose-Einstein
condensed and one can ignore mean field effects. 
The quantitative estimates of collisional effects will be
however based on classical statistics.

The force 
${\bf F}_{\rm trap} = - {\bf \nabla} U_{\rm trap}(x,y,z)$ is produced by
the
confining potential   which in the following will be chosen to be of 
harmonic form:
\begin{equation}
U_{\rm trap}(x,y,z) = 
\frac{1}{2}m\omega^{2}_\perp(x^2+y^2)+\frac{1}{2}m\omega^2_zz^2\;.
\label{potential}
\end{equation}
We introduce  the anisotropy parameter $\lambda=\omega_z/\omega_\perp$.
For $\lambda=1$, one deals with an isotropic harmonic trap, for 
$\lambda^2 \ll 1$ one has a cigar shaped trap and for $\lambda^2 \gg 1$
a disk shaped trap. 
Starting from (\ref{Boltzmann}), one can derive useful equations
for the average of a general dynamical quantity $\chi({\bf r},{\bf v})$: 
\begin{equation}
\frac{d\langle\chi\rangle}{dt}-\langle{\bf v}\cdot   \mbox{\boldmath
$\nabla$}_{\bf r}\chi\rangle -\langle\frac{{\bf F}_{\rm trap}}{m}\cdot 
\mbox{\boldmath $\nabla$}_{\bf v}\chi\rangle = \langle\chi I_{\rm
coll}\rangle\;,
\label{conservth}
\end{equation}
where the average is taken in both
position and velocity space: 
\begin{equation}
\langle\chi\rangle = {1\over N}\int d^3{r}\;d^3{v}\;f({\bf r},{\bf
v},t)\;\chi ({\bf r},{\bf v})\;,
\label{average}
\end{equation}
The quantity
$\langle\chi I_{\rm coll}\rangle$ can be written in the useful form  
\begin{eqnarray}
\langle\chi I_{\rm coll}\rangle = {1\over 4N}  \int d^3{r}\;
d^3{v}_1\; [\chi_1+\chi_2-\chi_{1^{\prime}}-\chi_{2^{\prime}}]\;I_{\rm
coll}(f)
\;,
\label{collterm}
\end{eqnarray}
as a consequence of  the invariance
properties
of the cross section  (we have put $\chi_i=\chi({\bf r},{\bf 
v}_i)$).
The collisional contribution (\ref{collterm}) is equal to zero 
if $\chi$  corresponds to a dynamic quantity conserved during the elastic 
collision. This happens if $\chi$ can be written in the form
\cite{Huang,FORD63}
\begin{equation} 
\chi = a({\bf r}) + {\bf b}({\bf r})\cdot {\bf v} +
c({\bf r}){\bf v}^2\;. 
\label{conservation} 
\end{equation}

\end{section}

\begin{section}{Monopole oscillation in harmonic isotropic traps}

Let us consider an harmonic isotropic trapping potential 
($\omega_x=\omega_y=\omega_z \equiv \omega_0)$. 
As a first application of (\ref{conservth}), one can immediately
derive the behaviour of 
the monopole mode\cite{FORD63,Cercignani,Boltzmann} by computing the
evolution 
of the square radius:
\begin{equation} 
\frac{d\langle {\bf r}^2\rangle}{dt}=2\langle {\bf r}.{\bf v}\rangle
\label{m1}
\end{equation}
In order to obtain a closed set of equations one also needs the following 
equations
\begin{equation} 
\frac{d\langle {\bf r}.{\bf v}\rangle}{dt}=\langle {\bf v}^2\rangle - 
\omega_0^2\langle {\bf r}^2 \rangle
\label{m2}
\end{equation}
and
\begin{equation} 
\frac{d\langle {\bf v}^2\rangle}{dt}=\omega_0^2\langle {\bf r}^2\rangle
\quad .
\label{m3}
\end{equation}
force and for the energy. 
The collisional term does not contribute to the
above equations because all the dynamic quantities 
satisfy the criterium (\ref{conservation}).
So there is no damping  for the ``breathing'' mode  of a
classical 
dilute gas confined in an  harmonic isotropic trap. 
The same is true if one includes quantum degeneracy effects in the 
collisional term of the Boltzmann equation.

By looking for solutions of eqs.(\ref{m1}-\ref{m3}) evolving in time 
as $e^{i\omega t}$ one immediately finds the result $\omega=2\omega_0$, 
holding for all collisional regimes from the collisionless to the 
hydrodynamical one. The occurrence of this monopole undamped solution was 
first pointed out by Boltzmann (see, for example, the discussion in
\cite{Cercignani}).

It is worth noticing that the frequency of the classical monopole
oscillation in isotropic harmonic traps differs from the 
one of a Bose-Einstein condensed gas at $T=0$. 
In the latter case the monopole oscillation 
is still undamped, but the frequency, for large $N$, 
is  $\omega=\sqrt5\, \omega_0$ \cite{hds}.
The difference is  the consequence of the  mean field interaction
which is absent in the Boltzmann equation (\ref{Boltzmann}).
Furthermore, at finite temperature
the monopole oscillation is expected to exhibit damping
because of the coupling between the condensate and the thermal
component of the gas.

\end{section}

\begin{section}{Quadrupole oscillation and relaxation time
approximation}

The purpose of this section is to investigate  the quadrupole mode of a 
classical gas as well as its coupling with
the monopole oscillation arising in  anisotropic traps.
In this case the solution of the Boltzmann equation exhibits damping and
one 
has to deal explicitly with the collisional term. 
In the presence of anistropy, the $l_z=0$ component of the
quadrupole is coupled with the monopole and one finally  finds the 
following set of coupled equations :
$$
\frac{d\langle\chi_1\rangle}{dt}-2\langle\chi_3\rangle=0
$$
$$
\frac{d\langle\chi_2\rangle}{dt}-2\langle\chi_4\rangle=0
$$
$$
\frac{d\langle\chi_3\rangle}{dt}-\langle\chi_5\rangle+
\frac{2\omega^2_\perp+\omega^2_z}{3}\langle\chi_1\rangle
+\frac{\omega^2_z-\omega^2_\perp}{3}\langle\chi_2\rangle
=0
$$
$$
\frac{d\langle\chi_4\rangle}{dt}-\langle\chi_6\rangle+
\frac{2\omega^2_z-2\omega^2_\perp}{3}\langle\chi_1\rangle+
\frac{\omega^2_\perp+2\omega^2_z}{3}\langle\chi_2\rangle
=0
$$
$$
\frac{d\langle\chi_5\rangle}{dt}+
\frac{2\omega^2_z+4\omega^2_\perp}{3}\langle\chi_3\rangle+
\frac{2\omega^2_z-2\omega^2_\perp}{3}\langle\chi_4\rangle
=0
$$
\begin{equation}
\frac{d\langle\chi_6\rangle}{dt}
+\frac{4\omega^2_z-4\omega^2_\perp}{3}\langle\chi_3\rangle+
\frac{4\omega^2_z+2\omega^2_\perp}{3}\langle\chi_4\rangle
=\langle\chi_6I_{\rm coll}\rangle
\label{systemQM}
\end{equation}
where we have defined the quantities
\begin{eqnarray}
\chi_1&=&{\bf r}^2
\nonumber\\
\chi_2&=&2z^2-r_\perp^2
\nonumber\\
\chi_3&=&{\bf r}\cdot{\bf v}
\nonumber\\
\chi_4&=&2zv_z-{\bf r}_\perp\cdot{\bf v}_\perp
\nonumber\\
\chi_5&=&{\bf v}^2
\nonumber\\
\chi_6&=&2v^2_z-v^2_\perp\;.
\end{eqnarray}
If the trap is isotropic ($\omega_{\perp}=\omega_z$),  (\ref{systemQM}) 
decouples in two subsystems. One subsystem refers  to the undamped monopole
oscillations discussed in the previous section. The other corresponds to
the damped quadrupole mode.
Notice that collisions affects only the last equation of (\ref{systemQM}).
Actually only
the variable $\chi_6=2v^2_z-v^2_\perp$ is not a conserved quantity
and hence does not satisfy the 
criterium (\ref{conservation}).
The collisional contribution  can be estimated by treating
the collisional integral in the so-called  relaxation time approximation 
\cite{FORD63}:
\begin{equation}
I_{\rm coll}(f)=-\frac{f-f_{\ell .  e.}}{\tau}
\label{le}
\end{equation}
where $\tau$ is a relaxation time. As we will see in the next section
 this time can be explicitly calculated in terms of the scattering
cross section.
The function 
$f_{\ell . e.}$ entering (\ref{le}) is the local equilibrium 
distribution function and is characterized by 
the same density, velocity field and local temperature as the 
distribution function $f$. 
Differently from $f$ it has, however, an isotropic velocity distribution.
As a consequence 
the average of $\chi_6$ on $f_{\ell . e.}$ exactly vanishes and one can
write
\begin{equation}
\langle\chi_6 I_{\rm coll}\rangle =-\frac{\langle\chi_6\rangle}{\tau}
\end{equation}
The system (\ref{systemQM}) is now linear, and one can search
for solutions of the type $e^{i\omega t}$ and deduce the associated
determinant. The result is: 
\begin{eqnarray}
 (\omega^2-4\omega_z^2)(\omega^2-4\omega_\perp^2)
-\nonumber\\
\frac{i}{\omega \tau}\left(\omega^4 - 
\frac{2}{3}\omega^2 (5\omega_\perp^2+4\omega_z^2) + 
8\omega_\perp^2 \omega_z^2\right)=0
\label{reldip}
\end{eqnarray}
The first term of (\ref{reldip}) corresponds to the dispersion law for 
the pure collisionless regime ($\omega_z\tau\longrightarrow\infty$). 
In this case the eigenfrequencies coincide with the ones predicted by 
the single particle harmonic oscillator hamiltonian: 
$\omega_{\rm CL}=2\,\omega_z$ and  $\omega_{\rm CL}=2\,\omega_\perp$.
On the contrary, the term  multiplying $1/\tau$
refers to the pure hydrodynamical regime ($\omega_z\tau\longrightarrow
0$).

For the spherical trap, one gets $\omega_{\rm HD}=\sqrt{2}\,\omega_0$
and $2\,\omega_0$
for the quadrupole and monopole modes respectively. For
a cigar shaped configuration ($\lambda^2 \ll 1$) the two hydrodynamic solutions
have the form 
$\omega_{\rm HD}=\sqrt{12/5}\;\omega_z$ and
$\sqrt{10/3}\;\omega_\perp$ \cite{Griffin}, while for a disk trap 
($\lambda^2 \gg 1$) one finds $\omega_{\rm HD}=\sqrt{8/3}\;\omega_z$ and
$\sqrt{3}\;\omega_\perp$.

Formula (\ref{reldip}), which provides the proper interpolation
between the collisionless and hydrodynamic regimes, can 
be  simplified in the  case of a spherical, cigar, and disk
shaped trap. In fact, the dispersion law   
 (\ref{reldip}) can be written in all these limiting cases in the useful
form:
\begin{equation}
\omega^2=\omega_{\rm CL}^2+\frac{\omega_{\rm HD}^2-\omega_{\rm CL}^2}
{1+i\omega\tilde{\tau}}\;,
\label{pethickfor}
\end{equation}
typical of relaxation phenomena \cite{Pethick,Landau}. The
time $\tilde{\tau}$ is related to $\tau$ by a simple numerical
factor. For example $\tilde{\tau}=\tau$  for the quadrupole mode in 
the spherical case, 
and $\tilde{\tau}=6\tau/5$ for the lowest mode of the cigar shaped 
configuration.
A relevant feature of (\ref{pethickfor}) is the presence of an
imaginary part, associated with the damping 
of oscillations. By writing the  frequency as $\omega
=\omega_r+i\Gamma$ one finds, assuming $\Gamma \ll \omega_r$,  
\begin{equation}
\Gamma \simeq\frac{\tilde{\tau}}{2}\frac{\omega_{\rm CL}^2-\omega_{\rm
HD}^2}{1+(\omega_r\tilde{\tau})^2}\;.
\label{damping}
\end{equation}
Notice that the damping depends crucially
on the difference between the frequencies calculated in the collisional
and 
hydrodynamic regimes and exactly vanishes when these frequencies 
coincide. This happens, for example,  
in the monopole case for isotropic trapping, as discussed in the
previous section\cite{dipole}.

In the hydrodynamic limit ($\omega_r\tilde{\tau}\to 0$) the
damping predicted by 
(\ref{damping}) takes the form
\begin{equation}
\Gamma_{\rm HD}\simeq\frac{\tilde{\tau}}{2}(\omega_{\rm CL}^2-\omega_{\rm
HD}^2)\;,
\label{gammaH}
\end{equation}
while in the opposite regime ($\tau\to\infty$) one gets
\begin{equation}
\Gamma_{\rm CL}\simeq
\frac{1}{2\omega_{\rm CL}^2\tilde{\tau}}(\omega_{\rm CL}^2
-\omega_{\rm HD}^2)\;.
\end{equation}
A maximum for $\Gamma$ is found at $\omega_r\tilde{\tau}\sim 1$, 
leading to $\Gamma \sim (\omega_{\rm CL}^2-\omega_{\rm HD}^2)/\omega_r$. 
Around this value, the approximation leading
to (\ref{damping}) is no longer accurate, and one should rather use
(\ref{reldip}) or (\ref{pethickfor}).
\end{section}

\begin{section}{Gaussian ansatz}

\noindent
One expects the relaxation time $\tau$ to be of the same order
of the inverse of the collision
rate $\gamma_{\rm coll}$  \cite{wu97}:
\begin{equation}
\gamma_{\rm coll}=\frac{n(0)v_{\rm th}\sigma_0}{2}
\label{gcoll}
\end{equation}
giving the number of collisions undergone by a given atom per unit of time. 
In (\ref{gcoll}) $v_{\rm th}=\sqrt{8\theta_0/\pi m}$ is the thermal velocity and
$n(0)$ the central density. 

An explicit evaluation of this link can be obtained by making
a gaussian ansatz for the distribution function, which allows 
for an analytic evaluation of the collisional contribution to the 
equations of motion. In fact, a natural assumption to describe the $l_z=0$ 
modes characterized by axial symmetry is given by the parametrization 
\begin{eqnarray}
f({\bf r},{\bf v},t)=N\bigg(\frac{m}{2\pi}\bigg)^3
\frac{\gamma_\perp\gamma_z^{1/2}}{\theta_{\perp}\theta_z^{1/2}}
\qquad\quad\nonumber\\
e^{-mU_\perp^2/2\theta_{\perp}}e^{-mU_z^2/2\theta_z}
e^{-m(\gamma_\perp r_\perp^2+\gamma_zz^2)/2}\;\;
\label{f}
\end{eqnarray}
where $r_\perp=\sqrt{x^2+y^2}$ and 
${\bf U}={\bf v}-\langle{\bf v}\rangle$.
Eq. (\ref{f}) provides a natural generalization of the local
equilibrium distribution, by introducing a deformation not only in
coordinate space (taken into account by the $\gamma$ parameters), but
also in velocity space.
The gaussian ansatz (\ref{f}) can be shown to describe exactly the
monopole and
quadrupole oscillations both in the hydrodynamic and
collisionless regimes. In the former case, the velocity distribution is
isotropic and hence $\theta_\perp=\theta_z$. For the collisionless case, an 
exact solution of the equations of motion can be found in the form (\ref{f}). 
However, in this case, $\theta_\perp$ is different from $\theta_z$,
corresponding to configurations far from local equilibrium.
Only in the presence of isotropic trapping and for the monopole
oscillation
the hydrodynamic and
collisionless solutions coincide. In this case the ansatz (\ref{f}),
with $\theta_\perp=\theta_z$ and $\gamma_\perp=\gamma_z$
provides an exact solution of the Boltzmann equation. 

In the limit of small oscillations around the equilibrium configuration, 
the temperature in axial and transverse axis can be expanded around
the equilibrium temperature $\theta_0$:
\begin{eqnarray}
\theta_\perp&=&\theta_0+\delta\theta_\perp
\nonumber\\
\theta_z&=&\theta_0+\delta\theta_z
\label{temperature}
\end{eqnarray}
One then finds:
\begin{equation}
\langle  \chi_6\rangle=\frac{2}{m}(\delta\theta_z-\delta\theta_\perp)
\label{ave6}
\end{equation}
showing that $\langle\chi_6\rangle$ is, as expected, directly sensitive
to the anisotropy of the velocity distribution.
The collisional contribution to the equation for $\chi_6$ (last eq. of
the system (\ref{systemQM})), can be also expressed in terms of this
anisotropy. 
By inserting eq. (\ref{f}) into
eq. (\ref{collterm})  with $\chi =\chi_6$ 
we obtain, after linearization, the expression
$$\langle\chi_6 I_{\rm coll}\rangle=\frac{1}{32\pi}
\frac{m\sigma_0}{N\theta_0^2}\int d^3{r} d^3{U}_1 d^3{U}_2
|{\bf U}_1-{\bf U}_2|d^2\Omega f_0(1)$$
$$f_0(2)\Delta\chi_6
\bigg[\delta\theta_{\perp}\bigg((U_{\perp}^2)_{1^{\prime}}+
(U_{\perp}^2)_{2^{\prime}}
-(U_{\perp}^2)_{1}-(U_{\perp}^2)_{2}\bigg)+$$
\begin{equation}
\delta\theta_z\bigg((U_{z}^2)_{1^{\prime}}+(U_{z}^2)_{2^{\prime}}
-(U_{z}^2)_{1}-(U_{z}^2)_{2}\bigg) \bigg]\;,
\label{CT}
\end{equation}
where $f_0$ is the gaussian (\ref{f}) evaluated at
equilibium, and 
$\Delta\chi_6=(\chi_6)_1+(\chi_6)_2-(\chi_6)_{1^{\prime}}-
(\chi_6)_{2^{\prime}}$. After some length but straightforward algebra
(see Appendix) the integral (\ref{CT}) can be written in the useful
form:
\begin{equation}
\langle \chi_6 I_{\rm coll}
\rangle=-\frac{2}{\tau}\frac{(\delta\theta_z-
\delta\theta_\perp)}{m}
=-\frac{\langle \chi_6\rangle}{\tau}\;,
\label{tg1}
\end{equation}
with $\tau$ given by
\begin{equation}
\tau=\frac{5}{4\gamma_{\rm coll}}\;.
\label{tg}
\end{equation}
Eq. (\ref{tg}) provides the explicit link between $\gamma_{\rm coll}$
and the relaxation time for the quadrupole 
$l_z=0$ mode. 
Notice that this relationship, provided by the gaussian ansatz (\ref{f}),
is independent of the deformation of the trap.
In the next section we will compare this prediction with the
result of a numerical calculation based on molecular dynamics.

\end{section}

\begin{section}{numerical simulation}

In this section, we  present results for the dispersion law arising from a 
numerical analysis, based on a molecular dynamics simulation.
We consider $N=2 \times 10^4$
particles moving in the potential (\ref{potential}). 
Binary elastic collisions are taken into account using
a boxing technique \cite{wu96,dgo98}. At each time step $\delta t$, the
position of each particle is discretized on a square lattice with a step 
$\xi$. The volume $\xi^3$ of a box is chosen such that the average
occupation 
$p_{\rm occ}$ of any box is much smaller than 1. 
Collisions occur only
between two particles occupying the same box, and  
the time step $\delta t$ is adjusted in such 
a way that the probability $p_{\rm coll}$ of a collisional event during 
$\delta t$ is also much smaller than 1. 
We choose typically  $p_{\rm occ}\sim p_{\rm coll}\sim  5\%$.

Initial conditions for exciting the lowest energetical mode are 
obtained
by a deformation in coordinate and velocity spaces of the cloud along 
the weak axes, keeping the phase space density 
constant. We have checked that this method leads to the excitation of
only
the lowest frequency mode. Then, we let the cloud evolve. 
The damped oscillation of the dynamical variable $\chi_2=2z^2-r_\perp^2$ is 
 analysed for different choices of the collision rate. 
As an example, the oscillation frequency and the damping
for a cigar shaped trap ($\lambda=1/10$) are 
plotted (solid circles) respectively on figures (1) and
(2). One observes that the frequency decreases as the ratio
$\omega_z/\gamma_{\rm coll}$
decreases and tends asymptotically to the hydrodynamic value 
$\sqrt{12/5}\;\omega_z=1.55\;\omega_z$. For large value of 
$\omega_z/\gamma_{\rm coll}$, the frquency instead approaches the collisionless
value $2\omega_z$. By performing a least square fit with formula 
(\ref{reldip}), we obtain:
\begin{equation}
\tau=(1.31 \pm 0.07)\frac{1}{\gamma_{\rm coll}}
\label{tautau}
\end{equation}
which well agree with the gaussian prediction (\ref{tg}).
The full line on figure (1) and (2) corresponds to eq. (\ref{reldip})
with $\tau$ given by the gaussian ansatz prediction (\ref{tg}).
We have checked that, for deformed cigar traps $(\lambda\ll 1)$, result
(\ref{tautau}) is independent of the value of $\lambda$.

We have instead found that, as the deformation parameter $\lambda$ becomes
larger, the results of the numerical simulation exhibits 
an increasing deviation from the gaussian prediction.
A discrepancy of the order of 25 \% has been observed for
$\lambda=\sqrt{8}$. The physical origin of this deviation remains to be 
undestood. In all cases, the gaussian ansatz underestimates the 
coefficient of proportionality between
the relaxation time $\tau$ and the inverse of the collision rate
$1/\gamma_{\rm coll}$.
\end{section}

\begin{section}{conclusion}
In this work, we have presented an investigation of the collective 
frequencies of a classical gas
trapped in an harmonic potential well. 
Starting from the classical Boltzmann equation, we have derived
a set of coupled equations for the averages of the relevant dynamic 
variables associated with the monopole and quadrupole modes. 
The collisional term has been treated in the time relaxation
approximation. The corresponding relaxation time was evaluated by a gaussian 
ansatz for the distribution function. 
The quality of the gaussian prediction has been checked numerically 
by a numerical simulation based on molecular dynamics.
\end{section}

\appendix

\section{Collisional integral}

\label{A}
The appendix is devoted to the explicit calculation of the collisional 
integral (\ref{CT}).
Let us introduce the center of mass velocity ${\bf C}$ and the relative
velocity before (${\bf V}$) and after (${\bf V'}$) collision : 
\begin{eqnarray}
{\bf U}_1&=&{\bf C}+{\bf V}/2 \nonumber\\
{\bf U}_2&=&{\bf C}-{\bf V}/2 \nonumber\\
{\bf U}_{1^{\prime}}&=&{\bf C}+{\bf V'}/2 \nonumber\\
{\bf U}_{2^{\prime}}&=&{\bf C}-{\bf V'}/2 
\label{a1}
\end{eqnarray}
The conservation of kinetic energy during an elastic
collision ensures
\begin{equation}
V^2=V'^2\;,
\label{a2}
\end{equation}
so that the collisional integral can be rewritten in the form
\begin{eqnarray}
\langle\chi_6I_{\rm coll}\rangle=-
(\delta\theta_z-\delta\theta_{\perp})
\frac{3}{128\pi}\frac{m\sigma_0}{N\theta_0^2}\nonumber\\
\int d^3{ r}\;d^3{V}\;d^3{C}\;d^2\Omega\;V
\;f_0(1)\,f_0(2)\,[V_z^2-V_{z^{\prime}}^2]^2\;.
\label{lin2}
\end{eqnarray}
Let us first calculate the angular integral :
\begin{equation}
I_{\Omega}\equiv\int d\Omega\big[ V_z^2-V'^2_z\big]^2
\end{equation}
We introduce a reference frame $\Re$
linked to ${\bf V}$, the vectors of the associated orthonormal basis 
beeing $(\hat{\bf a},\hat{\bf b},\hat{\bf c})$. 
Without loss of generality we choose
$\hat{\bf a}$ such that ${\bf V}=V\cdot\hat{\bf a}$, and the $z$ axis in
the plane
generated by $(\hat{\bf a},\hat{\bf b})$. 
The relative velocity ${\bf V}'$ is characterized in
$\Re$ by two spherical angles
$(\theta',\varphi')$ : 
\begin{eqnarray}
{\bf V}'.\hat{\bf a}&=&V\cos\theta' \\
{\bf V}'.\hat{\bf b}&=&V\sin\theta'\cos\varphi' \\
{\bf V}'.\hat{\bf c}&=&V\sin\theta'\sin\varphi'
\end{eqnarray}
Thus
$$
V'_z={\bf V}'\cdot\hat{\bf z}=V_z\cos\theta'+
V\sin\theta'\cos\varphi'(\hat{\bf b}\cdot\hat{\bf z})\;,
$$
where $\hat{\bf z}$ is the unit vector of the $z$ axis and
\begin{eqnarray*}
V'^2_z&=&V^2_z\cos^2\theta'+V^2\sin^2\theta'\cos^2\varphi'(\hat{\bf
b}\cdot
\hat{\bf z})^2\\
&+&2VV_z\cos\theta'\sin\theta'\cos\varphi'(\hat{\bf b}\cdot\hat{\bf
z})\;.
\end{eqnarray*}
With our choice of coordinate, one has
$$\hat{\bf z}\cdot\hat{\bf z}=1=(\hat{\bf a}\cdot\hat{\bf z})^2+
(\hat{\bf b}\cdot\hat{\bf z})^2\;,$$
which implies
$$
V^2(\hat{\bf b}\cdot\hat{\bf z})^2=V^2-V^2_z
$$
and, finally,
\begin{eqnarray*}
V^2_z-V'^2_z&=&V^2_z(1-\cos^2\theta')-\sin^2\theta'\cos^2\varphi'(V^2-V^2_z)\\
&2&VV_z\sin\theta'\cos\theta'\cos\varphi'(\hat{\bf b}\cdot\hat{\bf
z})\;.
\end{eqnarray*}

By integrating the square of the previous expression, one finds
\begin{equation}
I_\Omega=\frac{32\pi}{15}\left(\frac{15}{8}V^4_z+\frac{3}{8}V^4-
\frac{5}{4}V^2_zV^2\right)\;.
\label{Iomega}
\end{equation}
The calculation of
the collisional integral (\ref{lin2}) is now straightforward, and finally 
yields the result :
\begin{equation}
\langle\chi_6I_{\rm coll}\rangle=-
(\delta\theta_z-\delta\theta_{\perp})\frac{4}{5m}v_{\rm th}\sigma_0n(0)\;,
\label{collint}
\end{equation}
where $v_{\rm
th}=\sqrt{8\theta_0/\pi
  m}$ is the thermal velocity of a particle of the gas.
  (\ref{collint}) permits to derive (\ref{tg1},\ref{tg}) with
  $\gamma_{\rm coll}$ defined by eq. (\ref{gcoll}).

\begin{figure}
\begin{center}
\epsfig{file=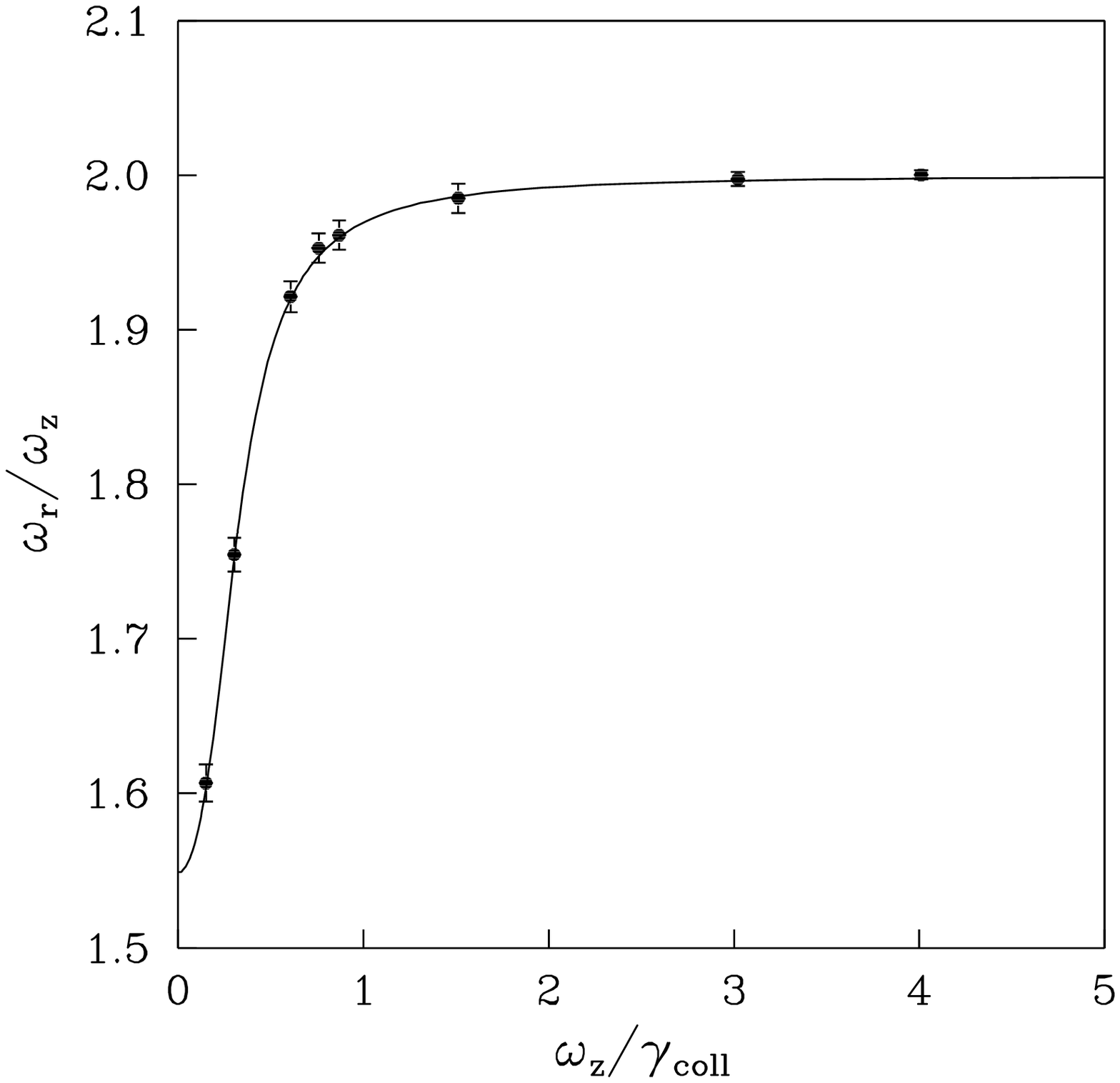, width=0.85\linewidth}
\begin{caption}
{Real part of frequency of the $l_z=0$ mode of a classical gas confined 
in a cigar shaped trap ($\lambda =1/10$), 
versus $\omega_z/\gamma_{\rm coll}$.  
The solid curve represents the prediction of the gaussian ansatz. 
The circles are the
numerical results obtained with a moleclular dynamics simulation.}
\end{caption}
\label{figReomega}
\end{center}
\end{figure}

\begin{figure}
\begin{center}
\epsfig{file=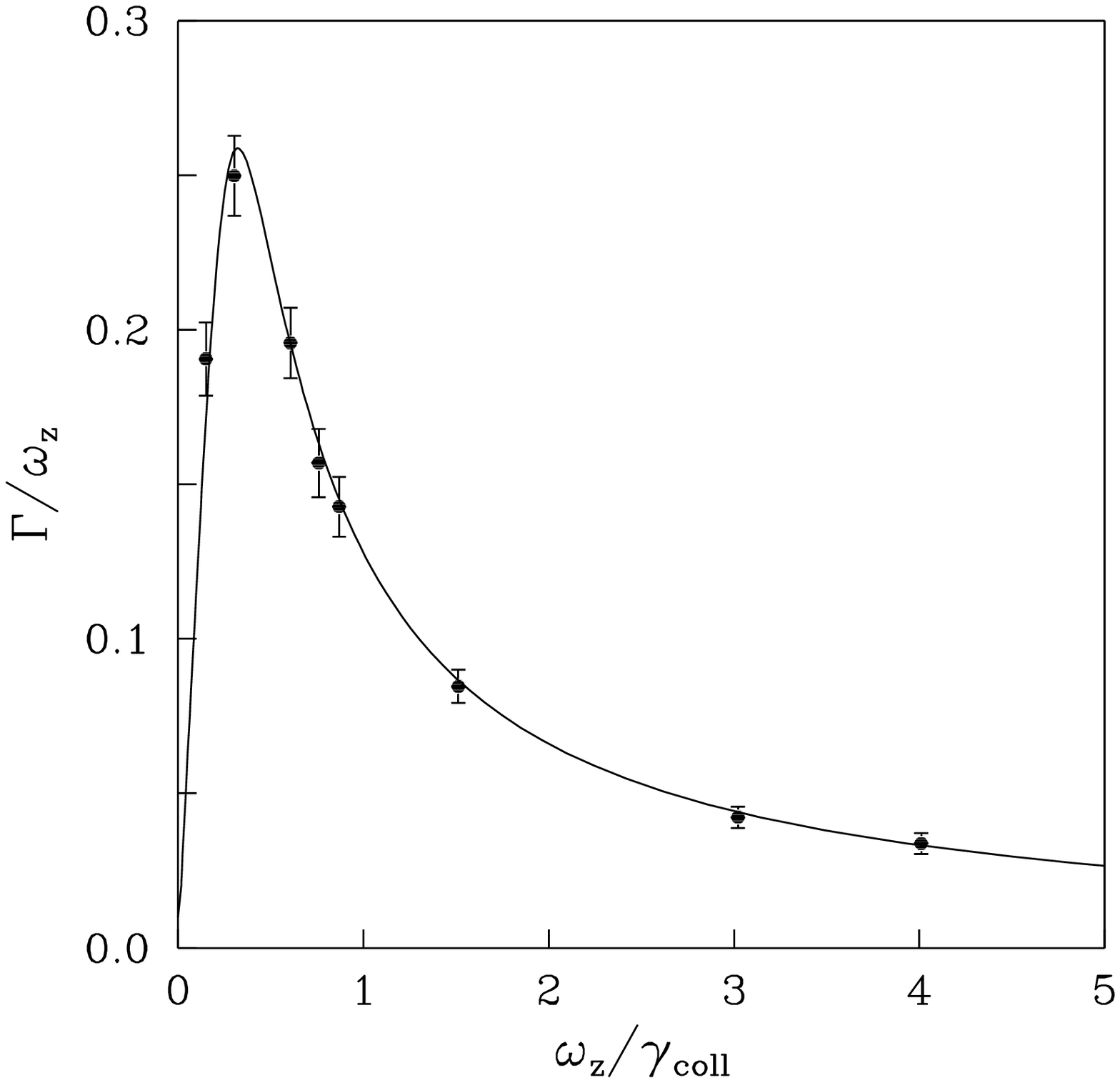, width=0.85\linewidth}
\begin{caption}
{Damping of the $l_z=0$ mode of a classical gas confined in a cigar
shaped
trap ($\lambda =1/10$), versus $\omega_z/\gamma_{\rm coll}$. 
The notations for the line and the markers are the same as
in figure 1.}
\end{caption}
\label{figdamping}
\end{center}
\end{figure}


\begin{references}
\bibitem{dipole}
Another important case where the damping identically vanishes
is the  center of mass (dipole) oscillation, whose frequency
coincides with the one of the harmonic potential, in both
hydrodynamic and collisionless regimes.
\end{references}
\end{document}